\def\preprint{1}                %  preprint
\def\comment#1{}
\if\preprint1
        \documentclass[useAMS,usenatbib]{mn2e}
        \usepackage{natbib}
        \usepackage{times}
      \usepackage[T1]{fontenc} 
        \usepackage{aecompl} 
        \usepackage{graphicx}
         \usepackage{url}
        \usepackage{calc}
        \usepackage{amssymb,amsmath}
        \usepackage{rotating}
         \usepackage{lscape}
         \bibpunct{(}{)}{;}{a}{}{,} % to follow the A&A style
         \usepackage{longtable}
          \usepackage{enumitem}
         \usepackage{float}
         \usepackage{subfigure}
       
\else
        \documentstyle[astrop-bib,referee,times]{mn2e}
        \newcommand{\includegraphics}[1]{}
        
\fi

\def\oversim#1#2{\lower0.5pt\vbox{\baselineskip0pt \lineskip-0.5pt
     \ialign{$\mathsurround0pt #1\hfil##\hfil$\crcr#2\crcr\sim\crcr}}}
\title[mm polarisation study of OH 231.8+4.2]{Millimeter polarisation of the protoplanetary nebula OH 231.8+4.2: A follow-up study with CARMA}
\author[L. Sabin et al.]{L. Sabin$^{1,2}$\thanks{E-mail:lsabin@astro.iam.udg.mx (LS)}, C.L.H. Hull$^{3}$\thanks{Jansky Fellow of the National Radio Astronomy Observatory, which is a facility of the National Science Foundation operated under cooperative agreement by Associated Universities, Inc.}, R.L. Plambeck$^{4}$, A.A. Zijlstra$^{5}$, R. V\'azquez$^{6}$, S.G. Navarro$^{1}$ \newauthor and P.F. Guill\'en$^{6}$ \\
$^{1}${Instituto de Astronom{\'i}a y Meteorolog{\'i}a, Departamento de F{\'i}sica, CUCEI, Universidad de Guadalajara, Av. Vallarta 2602, C.P. 44130, Guadalajara, Jal., Mexico}\\
$^{2}$ Instituto de Estudios Avanzados de Baja California, A. C., Av. Obreg\'on 1755, 22800 Ensenada, BC, Mexico.\\
$^{3}$ Harvard-Smithsonian Center for Astrophysics, 60 Garden Street, Cambridge, MA 02138, USA\\
$^{4}$Astronomy Department \&  Radio Astronomy Laboratory, University of California, Berkeley, CA 94720-3411, USA\\
$^{5}$Jodrell Bank Centre for Astrophysics, Alan Turing Building, University of Manchester, Manchester, M13 9PL, UK\\
$^{6}$Instituto de Astronom\'{i}a, Universidad Nacional Aut\'{o}noma de M\'{e}xico, Apdo. Postal 877, 22800 Ensenada, B. C, Mexico.\\
}
\begin{document}

\date{Accepted, 26 February 2015. Received, 26 February 2015; in original form 19 August 2014.}

\pagerange{\pageref{firstpage}--\pageref{lastpage}} \pubyear{2014}

\maketitle

\label{firstpage}

\begin{abstract}

 In order to investigate the characteristics and influence of the magnetic field in evolved stars, we performed a follow-up investigation of our previous submillimeter analysis of the proto-planetary nebula (PPN) OH 231.8+4.2 \citep{Sabin2014}, this time at 1.3\,mm with the CARMA facility in polarisation mode for the purpose of a multi-scale analysis. OH 231.8+4.2 was observed at $\sim$\,$2.5\arcsec$ resolution and we detected polarised emission above the 3$\sigma$ threshold (with a mean polarisation fraction of 3.5\%). The polarisation map indicates an overall organised magnetic field within the nebula. The main finding in this paper is the presence of a structure mostly compatible with an ordered toroidal component that is aligned with the PPN's dark lane. We also present some alternative magnetic field configuration to explain the structure observed. These data complete our previous SMA submillimeter data for a better investigation  and understanding of the magnetic field structure in OH 231.8+4.2.

\end{abstract}


\section{Introduction}

The search for magnetic fields (hereafter B-fields) in evolved low and intermediate mass stars, from Asymptotic Giant Branch (AGB) stars to Planetary Nebulae (PNe), has been boosted in recent years by observations obtained through multiple techniques. The detections of synchrotron emission \citep{Perez2013}, Faraday rotation \citep{Ransom2010}, maser emission \citep{Ferreira2013,Wolak2012,Amiri2011}, dust continuum polarisation \citep{Sabin2007}, and molecular line polarisation \citep{Girart2012,Vlemmings2012} have allowed us to determine the configuration and/or strength of the magnetic field \footnote{It is worth mentioning that in PNe, contrary to the case of AGB stars (see the work by \citealt{Lebre2014}), there is still no conspicuous and definite measurements indicating the presence of a B-field at the stellar surface \citep{Jordan2012,Leone2011,Leone2014,Steffen2014}.}. All these discoveries are leading to a better understanding of the role magnetic fields play in the dynamics of evolved stars---in particular, the role they play in shaping the circumstellar envelopes in PNe.

In the case of dust grain polarisation, the assumption made is that in the presence of a magnetic field, non-spherical spinning dust grains will be aligned with their long axis perpendicular to the B-field lines \citep{Lazarian2003,LazarianHoang2007}. The inferred magnetic field orientations are therefore obtained by rotating the polarisation vectors by 90$^{\circ}$ allowing us to trace the direction of the magnetic field projected onto the plane of the sky. The polarisation measurements at large wavelengths (i.e. from the far-infrared regime and above) are strongly correlated to the mass or intensity of the regions observed in the line of sight and thus the denser/heavier zones will have a larger weight on the determination of the polarisation characteristics.

Recently, our team \citep[hereafter Paper I]{Sabin2014} performed a new investigation of the dust continuum polarisation in the protoplanetary nebulae (PPNe) CRL 618 and OH 231.8+4.2 at 345\,GHz with the Submillimeter Array (SMA), an interferometer of eight 6m dishes. The array was used in compact configuration ($\sim$2$\arcsec$ resolution).

The main result of Paper I was the discovery in OH 231.8+4.2 of an ``X shaped'' B-field pattern consistent with a poloidal configuration. There were also hints of a toroidal B-field component in the SMA data. The alignment between the magnetic field structure and the $^{12}$CO($ J {=} 3 \rightarrow 2$) molecular outflow indicated a dynamically important field and the possible presence of a magnetic launching mechanism compatible with the magneto-centrifugal models of \citet{Blackman2001N} and \citet{Blackman2014}.

 In an effort to constrain these results we performed a follow-up investigation, this time at 1.3\,mm, in order to obtain a multi-scale and multi-wavelength coverage of this source. The new polarimetric observations provide more complete and spatially extended magnetic field maps that will allow better comparisons with magneto-hydrodynamical (MHD) models.

In $\S$2 we describe our observations and data reduction, in $\S$3 we present the dust continuum and polarisation results for OH 231.8+4.2. The discussion, which includes the comparison with the SMA data, and our conclusions are presented in $\S$4 and $\S$5, respectively. Finally, although this paper focused mainly on the source OH 231.8+4.2, we also chose to briefly present in the appendix section $\S$A the non-conclusive CARMA detection for the other PPN studied in Paper I, namely CRL 618.

\section{Observations and data reduction}

The observations presented in this article were carried out with the Combined Array for Research in Millimeter-wave Astronomy 
(CARMA, \citealt{Bock2006}). The 15-element array (six 10.4-meter and nine 6.1-meter antennas) was used in D-configuration, which had baseline lengths between 11\,m and 148\,m, and a resolution of $\sim$2.5$\arcsec$ at 230 GHz. The 1.3\,mm dual-polarisation receivers \citep{Hull2011} were tuned to include both 1.3\,mm continuum, the $^{12}$CO($ J = 2 \rightarrow 1$) spectral line at 230.538 GHz, and the SiO($J = 5 \rightarrow 4$) line at 217.105 GHz. We used a correlator configuration with six 500\,MHz wide bands to measure the dust continuum, and two 500\,MHz wide band to map spectral-line emission. Our observations were performed in full-Stokes mode, which allowed all four polarisation crossproducts ($LL$, $RR$, $LR$, $RL$) to be measured in order to derive maps of Stokes $I$, $Q$, $U$, and $V$. 

OH 231.8+4.2 was observed on 26 and 27 December 2013. The observing parameters for both sources are summarised in Table \ref{Configuration}.  It is worth noting the weather conditions during these track which was excellent for OH 231.8+4.2.

The gain, passband, and flux calibrations, as well as the full processing up to the map production, were performed with the software {\sc miriad} \citep{Wright1993,Sault2011}. The reduction of CARMA polarisation data required two additional steps: XYphase and leakage calibrations. The XYphase calibration corrects for the phase difference between the L- and R-circular receivers and is performed by observing a linearly polarised noise source with known position angle. The leakage calibration corrects any instrumental polarisation errors and is done by observing a strong source over a range of parallactic angles. 
 \citet[p.2]{Hull2013} and \citet[p.3]{Hull2014} derived in their survey TADPOL (and for the same configuration used here) a typical band-averaged leakage of around 6\%. More detailed information on the leakages can be found in these two references. The full description of polarisation observations with CARMA has been thoroughly discussed by \citet{Hull2014}.

\begin{figure}
\vspace{-0.8cm}
\begin{flushleft}
{\includegraphics[height=9cm]{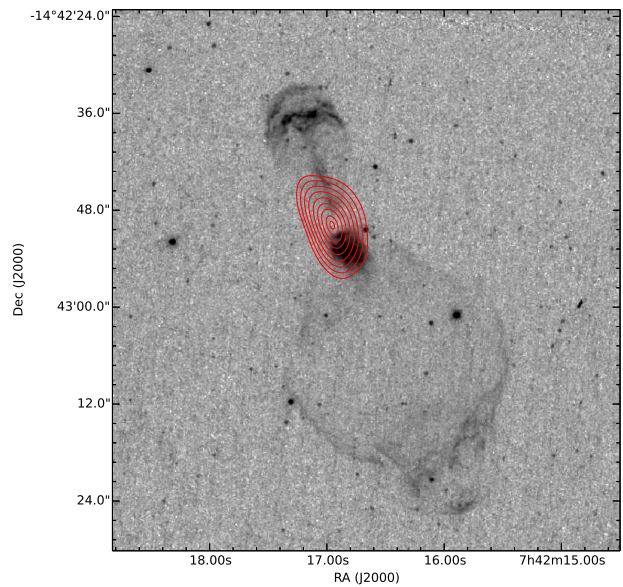}} %\hspace{2cm}
\caption{\label{OH231_cont} HST H$\alpha$ image of OH 231.8+4.2 (in logarithmic scale), with the superimposed CARMA 1.3\,mm dust continuum emission (red contours with values of (0.010, 0.022, 0.040, 0.065, 0.100, 0.140, 0.180, 0.220, 0.240) Jy/beam).}
\end{flushleft}
\end{figure}

\begin{table}
%\small\addtolength{\tabcolsep}{-4pt}
\caption[]{\label{Configuration} Summary of the CARMA observations}
\hspace{1cm}
\begin{flushleft}
\begin{tabular}{|l|l|l|}
\hline
Source & OH 231.8+4.2    &  \\
\hline
Phase center & $\alpha$=07:42:17.0 &  $\delta$=-14:42:50.2 \\
Obs. Date &  2013-12-27 & 2013-12-26 \\
Gain calibrator  & 0730-116 & 0730-116 \\
Passband calibrator & 3C84 & 3C84 \\
Flux calibrator & Mars  & Mars  \\
Total project length (hrs)  & 4.0 & 4.0 \\
Time on source (hrs) & 2.37 & 2.41 \\
Total Opacity$^{\dagger}$ & 0.19  & 0.16 \\
\hline
\end{tabular}
\begin{minipage}{10cm}
\hspace{-1cm}
\hspace{1cm}$\dagger$ Opacity at 230 GHz due to phase noise and atmospheric absorption
\end{minipage}
\end{flushleft}
\end{table}

\begin{figure*}
%\begin{flushright}
{\includegraphics[height=7.5cm]{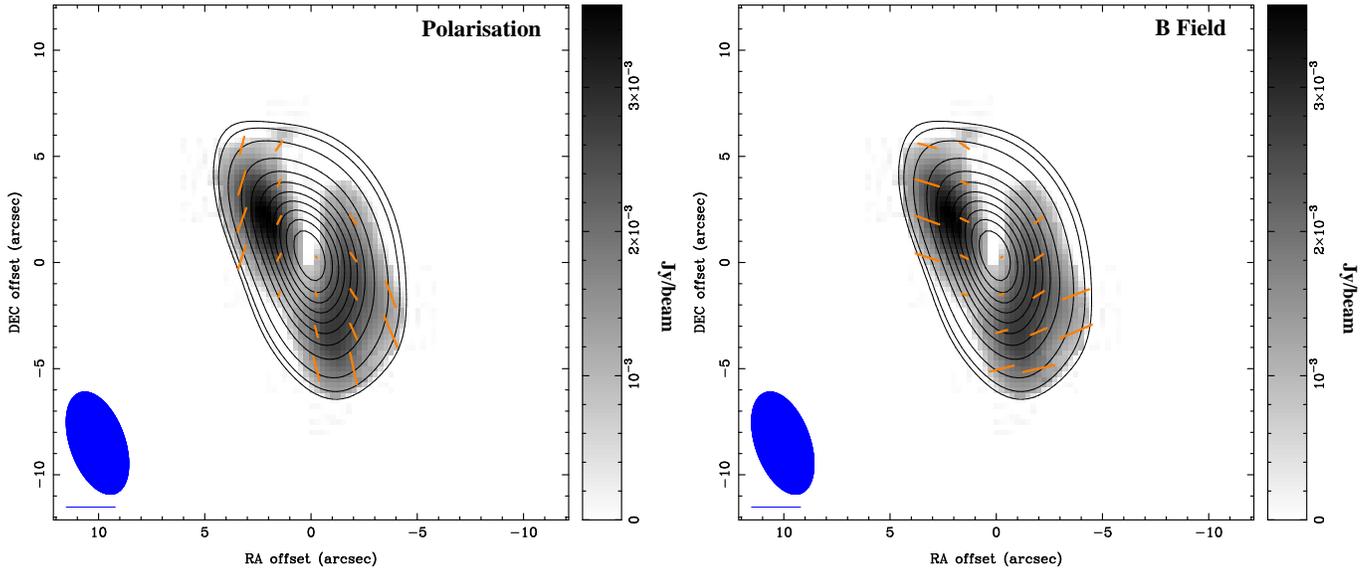}}
\caption{\label{polarisation} Left Panel: Millimeter polarisation map of OH 231.8+4.2. The black contours showing the total dust emission are drawn in steps of 0.0241 Jy/beam $\times$ (0.2, 0.3, 0.5, 1, 2, 3, 4, 5, 6, 7, 8, 9). The grey scale image indicates the polarised intensity in Jy/beam. Finally, the polarisation vectors are drawn as orange line segments; the scale (bottom left) is set to 10\%.  Right Panel: Inferred magnetic field map obtained by rotating the original polarisation vectors by 90$^{\circ}$. The map indicates a generally well organised magnetic field configuration within the whole structure. }
%\end{flushright}
\end{figure*}

\begin{figure}
%\begin{flushright}
{\includegraphics[height=7.5cm]{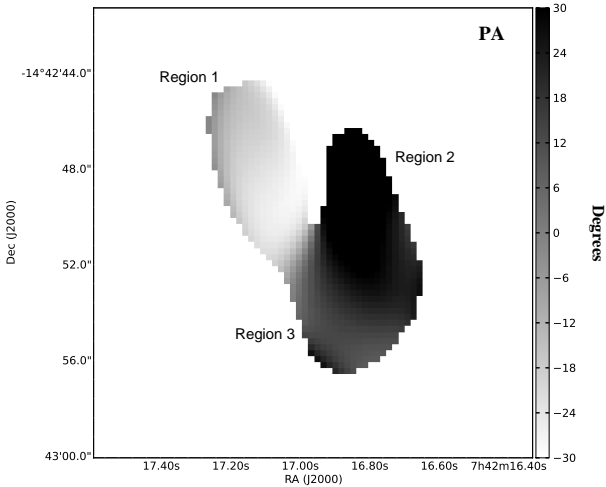}}
\caption{\label{PA} Zoomed-in view showing the variation of the polarisation position angle orientation. Three regions can be highlighted: region 1 with a mean PA of $-21 \pm 7^{\circ}$, region 2 with a mean PA of $+34 \pm 7 ^{\circ}$ and finally region 3  with a mean PA of $+14 \pm 4 ^{\circ}$. Note that we need to add 90$^{\circ}$ to those numbers to have the magnetic orientation angle. Combined with Fig.\ref{polarisation}-Right, we observe that while the first two zones are a reminiscent of the upper portion of the X-shape magnetic structure reported in Paper I, the third zone is a new and interesting element which fits with the ``dark lane'' or torus of OH 231.8+4.2 (see \S5).}
%\end{flushright}
\end{figure}

\section{Polarisation and magnetic field in OH 231.8+4.2}

\subsection{Continuum polarisation}

The final continuum map of OH 231.8+4.2 has a synthesised beam size of $\sim$\,$4.6 \times 2.4\arcsec$ with a position angle of 18$^{\circ}$.  The rms noise derived for the Stokes $I$, $Q$, and $U$ maps was $\sigma_{I}$ = 1.7 mJy/beam and $\sigma_{Q,U}$ = 0.4 mJy/beam. The 1.3\,mm continuum is elongated in the NE--SW direction (partially owing to the extended beam) and expands over $\sim$\,$15.3 \times 8.8$\,arcsec$^{2}$ (see Figure \ref{OH231_cont}). We detected a peak intensity of $\sim$240 mJy/beam centred on the coordinates $\alpha={\rm 07^{h} 42^{m} 16\fs950}$, $\delta=-14\degr42\arcmin49\farcs95$ and a mean flux of $\sim$70 mJy/beam. 

Contrary to the ``patchy'' structure described in Paper I, the polarised continuum of OH 231.8+4.2 shown in Figure \ref{polarisation}, appears more uniform and extended ($\sim$ 14.39$\times$7.86 arcsec$^{2}$). A comparison of both datasets is discussed in \S 4.1. We measured a peak polarised intensity ($p_\textrm{pk}$) of about 3.6 mJy/beam and a mean of $\sim$1.7 mJy/beam (4.25$\sigma$) over the whole polarised region. Based on the careful analysis of the polarised intensity, percentage polarisation, and position angle maps, the polarisation distribution could be separated in three regions within two main areas (Figure \ref{PA}) : \\

\vspace{-\topsep}
\begin{itemize}
\setlength{\parskip}{0pt}
 \setlength{\itemsep}{0pt plus 1pt}

\item First an ``upper distribution'' in the East (region 1 in Figure \ref{PA}) where we found the peak $p_\textrm{pk}$. The percentage polarisation ($p$) in this area reaches 12\%, with a mean of 3.4\%. Figure \ref{polarisation} shows that in this area the nine electric vectors (E-vectors) are well organised with a mean PA of $-21 \pm 7^{\circ}$.  %leading to magnetic orientations of mean PA $\simeq +69^{\circ}$. 

\item The second area, or ``lower distribution,'' in the West is more extended than the upper distribution but is similar in terms of $p$ (peak at 12.3\% and a mean of 3.5\%). The main difference lies in the E-vector distribution. Figure \ref{PA} shows that while the region 2 has a mean PA of $+34 \pm 7^{\circ}$ described by four E-vectors, region 3 has a mean PA of $+14 \pm 4^{\circ}$ described by seven E-vectors, indicating different overall directions. The magnetic field appears to be well organised within each region. We point out that in both regions the rather large $p_\textrm{pk}$ is related to the outer edge of the nebula. 
\end{itemize}

%\vspace{-\topsep}

 We emphasise however, that the demarcation made between regions 2 and 3 is likely affected by the weight/number of E-vectors present in each area. Therefore instead of a clear distinction in the field orientation, we could actually be seeing a smooth variation of the magnetic field across the whole southern lobe.\\

The drop of the fractional polarisation from the edges to the central area (near the total peak intensity), the so called ``polarisation hole,'' is also seen in our map. The causes inducing this depolarisation effect have yet to be formally established, but are generally associated to three main phenomena. The first is beam depolarisation, where unresolved structures in the plane of the sky are averaged across the observing beam. Second, the low or null percentage polarisation can be linked to the poor alignment of the grains (i.e. low grain alignment efficiency) in high density regions. Finally, the fluctuation of the magnetic field along the line of sight is also likely to induce a change in the percentage polarisation.

\subsection{Magnetic field vs. $^{12}$CO($J = 2 \rightarrow 1$) molecular outflow}

\begin{figure}
%\vspace{-0.3cm}
%\begin{flushleft}
\hspace{-1cm}
{\includegraphics[height=9.2cm]{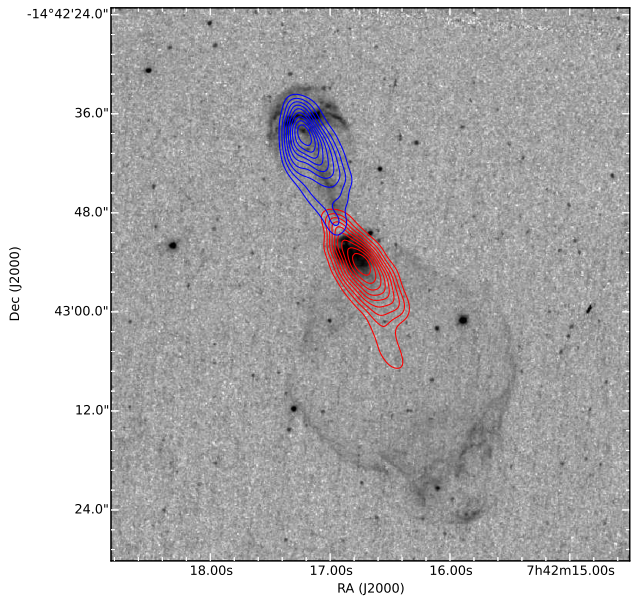}}
{\includegraphics[height=8.5cm,angle=-90]{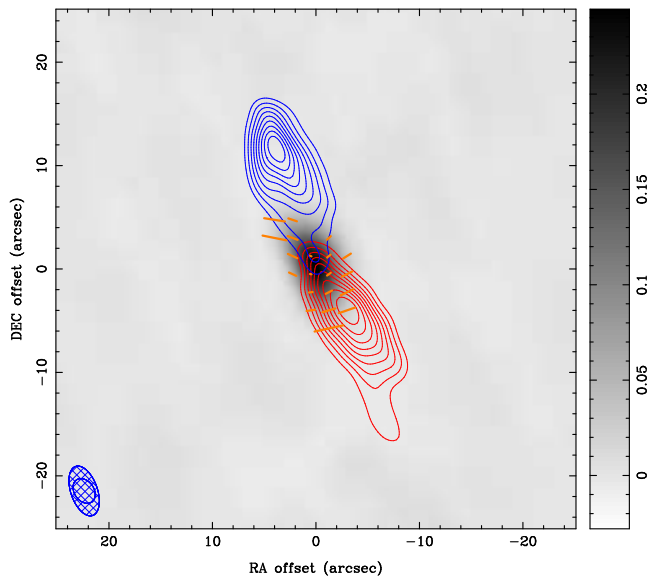}} %\hspace{2cm}
\caption{\label{outflow} Top: $^{12}$CO($J=2\rightarrow1$) molecular outflow detected with CARMA overlaid on the HST H$\alpha$ map. The blue shifted lobe is shown as blue contours in steps of 8.137 Jy/beam $\cdot$ km/s $\times$ (--5, --4, --3, --2, --1, 1, 2, 3, 4, 5, 6, 7, 8, 9); the red shifted  lobe is shown as red contours in step of 5.586 Jy/beam\,$\cdot$\,km/s $\times$ (--5, --4, --3, --2, --1, 1, 2, 3, 4, 5, 6, 7, 8, 9). Bottom: Map of the magnetic field ``vectors'' superimposed on the $^{12}$CO($J = 2 \rightarrow 1$) molecular outflow. The grey scale image traces the dust continuum emission.  The best observed pattern is the globally perpendicular alignment of the southern vectors with the red-shifted outflow.}
%\end{flushleft}
\end{figure}

\begin{figure}
\hspace{-1cm}
%\begin{flushleft}
{\includegraphics[height=9.2cm]{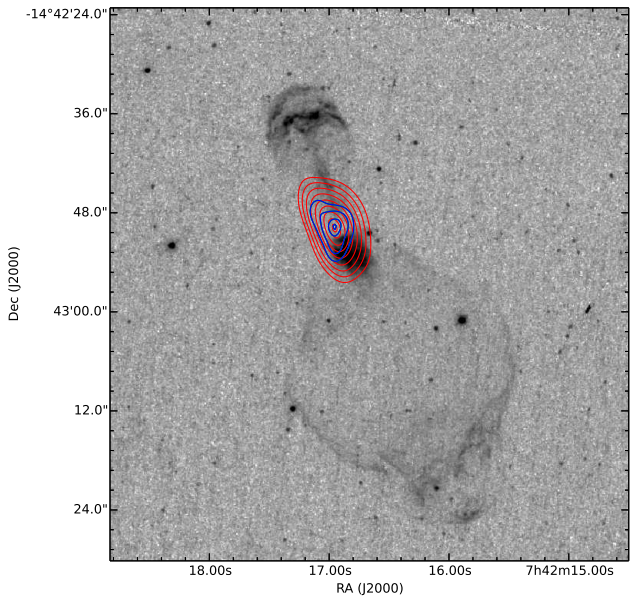}} %\vspace{-0.8cm}
\caption{\label{OH231_cont2} Distribution of the thermal dust continuum emission at $\sim$345 GHz in blue (inner distribution--SMA data) and $\sim$230 GHz in red (outer distribution--CARMA data), overlaid on optical H$\alpha$ data (HST). The combination of high resolution data will allow us to perform accurate multi-scale magnetic analysis of the PPN (see $\S$4).}
%\end{flushleft}
\end{figure}

The establishment of a correlation between the distribution of magnetic fields and outflow orientations would help determine whether PPN outflows are magnetically launched. Indeed, the conditions for such launching mechanisms have been explained by \citet{Blackman2001N}, and rely on the presence of a poloidal field at small distance from the central engine and a toroidal field further away (see Paper I). However, we are still hindered by a lack of information on the strength of the magnetic field.\\
\indent The most direct path to establish this correlation Bfield/outflow is to study the line polarisation (and trace the magnetic field) assuming the presence of a Goldreich-Kylafis effect \citep{Goldreich1981,Goldreich1982} which predicts that the emission from rotating molecules can be polarised in the presence of a magnetic field. The strong $^{12}$CO($J = 2 \rightarrow 1$) line in OH 231.8+4.2  (see also \citealt{Alcolea2001}) is therefore a good candidate to test the relationship between the magnetic field and the outflows. Unfortunately, the line polarisation mode is not available yet on CARMA. Therefore, as we did in Paper I, we rely on the comparison between the magnetic field derived from the continuum polarised emission and the direction and distribution of the CO emission line.

 Figure \ref{outflow} (top panel) shows the superposition of the CO emission on the H$\alpha$ HST image and we can immediately see the good correspondence between the blue-shifted emission and the northern optical outflow while the red-shifted CO outflow, roughly similar in size to its blue counterpart, only traces the interior (inner side) of the larger southern ionised outflow. In terms of polarisation, the same Figure \ref{outflow} (bottom panel), tends to indicate that the   orientation of the ``magnetic vectors'' in the region 3 from Figure \ref{PA} is globally roughly perpendicular to the red-shifted (lower) CO outflow. The low number of ``magnetic vectors'' corresponding to the blue-shifted (upper) CO lobe prevents us from performing any more precise correlation between the magnetic field and the outflow direction.

\section{Discussion}

%While little can be said about CRL 618 except for the confirmation of its low percentage polarisation, which is most likely linked to the dust grain nature and size (i.e. small carbonaceous grains, see $\S1$),
The new CARMA millimeter data help to complete our understanding of the magnetic field configuration in OH 231.8+4.2. 

\subsection{CARMA/SMA comparative study}

 The 1.3\,mm data led to more extended maps of continuum intensity and polarisation. This allows us to perform a two-scale comparison of the magnetic structure (Figure \ref{OH231_cont2}).
We find that the global mean continuum percentage polarisation of 3.4\% is slightly lower than the 4.6\% found with the SMA at $\sim$345 GHz. 
%In terms of configuration, OH 231.8+4.2 displays a northern V-shaped magnetic field structure which is consistent with the northern section of the X-shaped distribution seen in paper I, and which also encompasses the $^{12}$CO($J = 2 \rightarrow 1$) outflow. 

 In terms of configuration, the most interesting finding of this study is the detection of a group of organised ``magnetic field vectors'' globally perpendicular to both the ionised optical and molecular outflows; these orientations would then be coincident with \textit{an organised toroidal magnetic field}. The presence of this structure is important for understanding how the magnetic field helps channel the outflow and how it affects the dynamics of OH 231.8+4.2. \\ 
\indent 
By comparing the information of both the CARMA (with a maximum baseline of $\simeq$148m) and SMA (with a maximum baseline of $\simeq$77m) datasets and the respective polarisation maps, we observe that a toroidal magnetic structure, roughly aligned with the PPN equatorial plane, is (only) seen at larger scales. In Paper I we also suggested the presence of a toroidal magnetic structure, but the small set of vectors on which we relied (in the south-west area in the SMA data) were unlikely to belong to such configuration as the resulting torus would have an inclination in disagreement with the equatorial plane. Those vectors seen with the SMA are more likely to be part of the dipole field where the magnetic field lines are bending/curving. We also detected with CARMA what seems to be at first sight a separated poloidal component (north-east side) which could also be associated to a similar poloidal structure observed with the SMA in the northern lobes. However, the fact that those vectors are ``only'' seen on the eastern side combined with the possible variation of the vectors distribution in the South, stated in $\S$3.1, might also suggest the presence of a single configuration in the form of a helical magnetic field, instead of two distinct field structures\footnote{We note that in any case the helical configuration would be the combination of both a poloidal and toroidal configurations.}. Proper modelling will be necessary to unveil the correct magnetic configuration.\\

 The detected magnetic configurations (and the scales on which they were detected) would be a first step in probing the presence of a magnetic launching mechanism in OH 231.8+4.2, as described by \citet{Blackman2001N} regarding proto-planetary nebulae. Further work, mostly modelling, is needed to fully confirm this phenomenon.

\subsection{Combined views of the magnetic field in OH 231.8+4.2}

OH 231.8+4.2 has been the subject of many polarisation studies. We can cite first the work by \citet{Etoka2009}, who investigated the configuration of the 1667\,MHz OH maser emission and showed the presence a magnetic field aligned with the outflow at a radius of $\sim$2$\arcsec$ from the central star (matching the SMA scale, \citealt{Sabin2014}).

Then the high resolution Very Long Baseline Array (VLBA) observations by \citet{Ferreira2012}, with a synthesised beam size of $\sim$1.7 $\times$ $\sim$0.9 mas, allowed the detection of 30 H$_{2}$O masers. Only three spots exhibited linear polarisation, ranging from p=0.28\%  to p=1.15\%. Circular polarisation was clearly detected in only one of the brightest maser spots, leading to a value of $B_{||} = 44 \pm 7$\,mG. In this case no configuration of the field could be accurately determined. The authors derived a surface stellar magnetic field of $\sim$2.5\,G assuming a toroidal configuration. But if we assume that while going closer to the central star the field evolved from a mostly toroidal to a poloidal or dipole configuration (as our new results seems to suggest, but still keeping in mind the helical field hypothesis) this would then bring the stellar magnetic field strength to 140\,G up to more than 8\,kG respectively (L. Ferreira, private communication). A deeper magnetic map, would help confirm this result.

\section{Conclusion} 

 We present a study of the dust polarisation in OH 231.8+4.2 at 1.3 mm. The investigation follows up on a previous study conducted by our group at submillimeter range, and allows us to obtain a multi-scale map of the magnetic field distribution. In the case of OH 231.8+4.2, we successfully observed a globally organised magnetic field pattern in the form of a north-east poloidal section and what appears to be a toroidal configuration well distributed in the south. The presence of two distinct field configurations in the CARMA data is hampered by the geometry of the so-called poloidal field which is only coincident with one side of the outflow. An alternative explanation is the presence of a single configuration with a helical distribution including both the single-sided poloidal and the toroidal structures.\\
\indent 
The components found with CARMA complete those identified with the SMA at smaller scales; thus, we conclude that the magnetic structure of OH 231.8+4.2 would consist of an inner poloidal field that is surrounded by an outer toroidal or possibly helical field . 
Further MHD modelling will be needed to properly describe the magnetic field configuration in OH 231.8+4.2 and link it (or not) to the signature of a magnetic collimation and launching mechanism.\\
\indent 
Much smaller scales (down to 0.5$\arcsec$) can now be reached with ALMA with its newly available polarisation system. The high resolution ALMA data will allow us to complete our multi-scale investigation of the magnetic field closer to the central engine, and to establish the role of the field in the late stages of stellar evolution. \\

\section*{Acknowledgements}
We would like to thank the referee for his/her valuable comments which helped us improve the quality of this paper.
LS is supported by the CONACYT grant CB-2011-01-0168078 and this research was also partially supported by the PIFI-2014 program from the University of Guadalajara (Mexico). C.L.H.H. acknowledges support from an NSF Graduate Fellowship and from a Ford Foundation Dissertation Fellowship. RV is supported by PAPIIT-DGAPA-UNAM grant 107914. LS also thank Marco G\'omez for his help. Support for CARMA construction was derived from the Gordon and Betty Moore Foundation, the Kenneth T. and Eileen L. Norris Foundation, the James S. McDonnell Foundation, the Associates of the California Institute of Technology, the University of Chicago, the states of California, Illinois, and Maryland, and the National Science Foundation. Ongoing CARMA development and operations are supported by the National Science Foundation under a cooperative agreement, and by the CARMA partner universities. This research has also made use of the SIMBAD database, operated at CDS, Strasbourg, France and the NASA's Astrophysics Data System. We acknowledge the use of observations made with the NASA/ESA Hubble Space Telescope, obtained from the data archive at the Space Telescope Science Institute. STScI is operated by the Association of Universities for Research in Astronomy, Inc. under NASA contract NAS 5-26555.

\bibliographystyle{mn2e}

\bibliography{Sabin_CARMA}

\begin{thebibliography}{}

\bibitem[\protect\citeauthoryear{{Alcolea}, {Bujarrabal}, {S{\'a}nchez
  Contreras}, {Neri} \& {Zweigle}}{{Alcolea} et~al.}{2001}]{Alcolea2001}
{Alcolea} J.,  {Bujarrabal} V.,  {S{\'a}nchez Contreras} C.,  {Neri} R.,
  {Zweigle} J.,  2001, \aap, 373, 932

\bibitem[\protect\citeauthoryear{{Amiri}, {Vlemmings} \& {van
  Langevelde}}{{Amiri} et~al.}{2011}]{Amiri2011}
{Amiri} N.,  {Vlemmings} W.,    {van Langevelde} H.~J.,  2011, \aap, 532, A149

\bibitem[\protect\citeauthoryear{{Blackman}}{{Blackman}}{2014}]{Blackman2014}
{Blackman} E.,  2014, in Asymmetrical Planetary Nebulae VI conference,
  Proceedings of the conference held 4-8 November, 2013. Edited by C. Morisset,
  G. Delgado-Inglada and S. Torres-Peimbert {Constraining Engine Paradigms of
  Pre-Planetary Nebulae Using Kinematic Properties of their Outflows}

\bibitem[\protect\citeauthoryear{{Blackman}, {Frank}, {Markiel}, {Thomas} \&
  {Van Horn}}{{Blackman} et~al.}{2001}]{Blackman2001N}
{Blackman} E.~G.,  {Frank} A.,  {Markiel} J.~A.,  {Thomas} J.~H.,    {Van Horn}
  H.~M.,  2001, \nat, 409, 485

\bibitem[\protect\citeauthoryear{{Bock}}{{Bock}}{2006}]{Bock2006}
{Bock} D.~C.-J.,  2006, in {Backer} D.~C.,  {Moran} J.~M.,   {Turner} J.~L.,
  eds, Revealing the Molecular Universe: One Antenna is Never Enough Vol.~356
  of Astronomical Society of the Pacific Conference Series, {CARMA: Combined
  Array for Research in Millimeter-Wave Astronomy}.
p.~17

\bibitem[\protect\citeauthoryear{{Etoka}, {Zijlstra}, {Richards}, {Matsuura} \&
  {Lagadec}}{{Etoka} et~al.}{2009}]{Etoka2009}
{Etoka} S.,  {Zijlstra} A.,  {Richards} A.~M.,  {Matsuura} M.,    {Lagadec} E.,
   2009, in {Soonthornthum} B.,  {Komonjinda} S.,  {Cheng} K.~S.,   {Leung}
  K.~C.,  eds, The Eighth Pacific Rim Conference on Stellar Astrophysics: A
  Tribute to Kam-Ching Leung Vol.~404 of Astronomical Society of the Pacific
  Conference Series, {The Geometrical and Magnetic Structure of the
  Proto-Planetary Nebula OH 231.8+4.2 Traced by OH Maser Emission}.
p.~311

\bibitem[\protect\citeauthoryear{{Girart}, {Patel}, {Vlemmings} \&
  {Rao}}{{Girart} et~al.}{2012}]{Girart2012}
{Girart} J.~M.,  {Patel} N.,  {Vlemmings} W.~H.~T.,    {Rao} R.,  2012, \apjl,
  751, L20

\bibitem[\protect\citeauthoryear{{Goldreich} \& {Kylafis}}{{Goldreich} \&
  {Kylafis}}{1981}]{Goldreich1981}
{Goldreich} P.,  {Kylafis} N.~D.,  1981, \apjl, 243, L75

\bibitem[\protect\citeauthoryear{{Goldreich} \& {Kylafis}}{{Goldreich} \&
  {Kylafis}}{1982}]{Goldreich1982}
{Goldreich} P.,  {Kylafis} N.~D.,  1982, \apj, 253, 606

\bibitem[\protect\citeauthoryear{Hull, Plambeck \& Engargiola}{Hull
  et~al.}{2011}]{Hull2011}
Hull C.,  Plambeck R.,    Engargiola G.,  2011, in General Assembly and
  Scientific Symposium, 2011 XXXth URSI 1 mm dual-polarization science with
  carma.
IEEE, Istanbul, Turkey, pp~1--4

\bibitem[\protect\citeauthoryear{{Hull}, {Plambeck}, {Bolatto}, {Bower},
  {Carpenter}, {Crutcher}, {Fiege}, {Franzmann} \& {Hakobian}}{{Hull}
  et~al.}{2013}]{Hull2013}
{Hull} C.~L.~H.,  {Plambeck} R.~L.,  {Bolatto} A.~D.,  {Bower} G.~C.,
  {Carpenter} J.~M.,  {Crutcher} R.~M.,  {Fiege} J.~D.,  {Franzmann} E.,
  {Hakobian} N.~S.,  2013, \apj, 768, 159

\bibitem[\protect\citeauthoryear{{Hull}, {Plambeck}, {Kwon}, {Bower},
  {Carpenter}, {Crutcher}, {Fiege}, {Franzmann} \& {Hakobian}}{{Hull}
  et~al.}{2014}]{Hull2014}
{Hull} C.~L.~H.,  {Plambeck} R.~L.,  {Kwon} W.,  {Bower} G.~C.,  {Carpenter}
  J.~M.,  {Crutcher} R.~M.,  {Fiege} J.~D.,  {Franzmann} E.,    {Hakobian}
  N.~S.,  2014, \apjs, 213, 13

\bibitem[\protect\citeauthoryear{{Jordan}, {Bagnulo}, {Werner} \&
  {O'Toole}}{{Jordan} et~al.}{2012}]{Jordan2012}
{Jordan} S.,  {Bagnulo} S.,  {Werner} K.,    {O'Toole} S.~J.,  2012, \aap, 542,
  A64

\bibitem[\protect\citeauthoryear{{Lazarian}}{{Lazarian}}{2003}]{Lazarian2003}
{Lazarian} A.,  2003, \jqsrt, 79, 881

\bibitem[\protect\citeauthoryear{{Lazarian} \& {Hoang}}{{Lazarian} \&
  {Hoang}}{2007}]{LazarianHoang2007}
{Lazarian} A.,  {Hoang} T.,  2007, \mnras, 378, 910

\bibitem[\protect\citeauthoryear{{Leal-Ferreira}, {Vlemmings}, {Diamond},
  {Kemball}, {Amiri} \& {Desmurs}}{{Leal-Ferreira} et~al.}{2012}]{Ferreira2012}
{Leal-Ferreira} M.~L.,  {Vlemmings} W.~H.~T.,  {Diamond} P.~J.,  {Kemball} A.,
  {Amiri} N.,    {Desmurs} J.-F.,  2012, \aap, 540, A42

\bibitem[\protect\citeauthoryear{{Leal-Ferreira}, {Vlemmings}, {Kemball} \&
  {Amiri}}{{Leal-Ferreira} et~al.}{2013}]{Ferreira2013}
{Leal-Ferreira} M.~L.,  {Vlemmings} W.~H.~T.,  {Kemball} A.,    {Amiri} N.,
  2013, \aap, 554, A134

\bibitem[\protect\citeauthoryear{{L{\`e}bre}, {Auri{\`e}re}, {Fabas}, {Gillet},
  {Herpin}, {Konstantinova-Antova} \& {Petit}}{{L{\`e}bre}
  et~al.}{2014}]{Lebre2014}
{L{\`e}bre} A.,  {Auri{\`e}re} M.,  {Fabas} N.,  {Gillet} D.,  {Herpin} F.,
  {Konstantinova-Antova} R.,    {Petit} P.,  2014, \aap, 561, A85

\bibitem[\protect\citeauthoryear{{Leone}, {Corradi}, {Mart{\'{\i}}nez
  Gonz{\'a}lez}, {Asensio Ramos} \& {Manso Sainz}}{{Leone}
  et~al.}{2014}]{Leone2014}
{Leone} F.,  {Corradi} R.~L.~M.,  {Mart{\'{\i}}nez Gonz{\'a}lez} M.~J.,
  {Asensio Ramos} A.,    {Manso Sainz} R.,  2014, \aap, 563, A43

\bibitem[\protect\citeauthoryear{{Leone}, {Mart{\'{\i}}nez Gonz{\'a}lez},
  {Corradi}, {Privitera} \& {Manso Sainz}}{{Leone} et~al.}{2011}]{Leone2011}
{Leone} F.,  {Mart{\'{\i}}nez Gonz{\'a}lez} M.~J.,  {Corradi} R.~L.~M.,
  {Privitera} G.,    {Manso Sainz} R.,  2011, \apjl, 731, L33

\bibitem[\protect\citeauthoryear{{P{\'e}rez-S{\'a}nchez}, {Vlemmings}, {Tafoya}
  \& {Chapman}}{{P{\'e}rez-S{\'a}nchez} et~al.}{2013}]{Perez2013}
{P{\'e}rez-S{\'a}nchez} A.~F.,  {Vlemmings} W.~H.~T.,  {Tafoya} D.,
  {Chapman} J.~M.,  2013, \mnras, 436, L79

\bibitem[\protect\citeauthoryear{{Ransom}, {Kothes}, {Wolleben} \&
  {Landecker}}{{Ransom} et~al.}{2010}]{Ransom2010}
{Ransom} R.~R.,  {Kothes} R.,  {Wolleben} M.,    {Landecker} T.~L.,  2010,
  \apj, 724, 946

\bibitem[\protect\citeauthoryear{{Sabin}, {Zhang}, {Zijlstra}, {Patel},
  {V{\'a}zquez}, {Zauderer}, {Contreras} \& {Guill{\'e}n}}{{Sabin}
  et~al.}{2014}]{Sabin2014}
{Sabin} L.,  {Zhang} Q.,  {Zijlstra} A.~A.,  {Patel} N.~A.,  {V{\'a}zquez} R.,
  {Zauderer} B.~A.,  {Contreras} M.~E.,    {Guill{\'e}n} P.~F.,  2014, \mnras,
  438, 1794

\bibitem[\protect\citeauthoryear{{Sabin}, {Zijlstra} \& {Greaves}}{{Sabin}
  et~al.}{2007}]{Sabin2007}
{Sabin} L.,  {Zijlstra} A.~A.,    {Greaves} J.~S.,  2007, \mnras, 376, 378

\bibitem[\protect\citeauthoryear{{Sault}, {Teuben} \& {Wright}}{{Sault}
  et~al.}{2011}]{Sault2011}
{Sault} R.~J.,  {Teuben} P.~J.,    {Wright} M.~C.~H.,  2011, Astrophysics
  Source Code Library, p.~6007

\bibitem[\protect\citeauthoryear{{Steffen}, {Hubrig}, {Todt}, {Sch{\"o}ller},
  {Hamann}, {Sandin} \& {Sch{\"o}nberner}}{{Steffen}
  et~al.}{2014}]{Steffen2014}
{Steffen} M.,  {Hubrig} S.,  {Todt} H.,  {Sch{\"o}ller} M.,  {Hamann} W.-R.,
  {Sandin} C.,    {Sch{\"o}nberner} D.,  2014, ArXiv e-prints

\bibitem[\protect\citeauthoryear{{Stephens}, {Looney}, {Kwon}, {Hull},
  {Plambeck}, {Crutcher}, {Chapman}, {Novak}, {Davidson}, {Vaillancourt},
  {Shinnaga} \& {Matthews}}{{Stephens} et~al.}{2013}]{Stephens2013}
{Stephens} I.~W.,  {Looney} L.~W.,  {Kwon} W.,  {Hull} C.~L.~H.,  {Plambeck}
  R.~L.,  {Crutcher} R.~M.,  {Chapman} N.,  {Novak} G.,  {Davidson} J.,
  {Vaillancourt} J.~E.,  {Shinnaga} H.,    {Matthews} T.,  2013, \apjl, 769,
  L15

\bibitem[\protect\citeauthoryear{{Vlemmings}, {Ramstedt}, {Rao} \&
  {Maercker}}{{Vlemmings} et~al.}{2012}]{Vlemmings2012}
{Vlemmings} W.~H.~T.,  {Ramstedt} S.,  {Rao} R.,    {Maercker} M.,  2012, \aap,
  540, L3

\bibitem[\protect\citeauthoryear{{Wolak}, {Szymczak} \& {G{\'e}rard}}{{Wolak}
  et~al.}{2012}]{Wolak2012}
{Wolak} P.,  {Szymczak} M.,    {G{\'e}rard} E.,  2012, \aap, 537, A5

\bibitem[\protect\citeauthoryear{{Wright} \& {Sault}}{{Wright} \&
  {Sault}}{1993}]{Wright1993}
{Wright} M.~C.~H.,  {Sault} R.~J.,  1993, \apj, 402, 546

\end{thebibliography}

\newpage
\appendix

\section{The particular case of CRL 618}
While this paper focused on the source OH 231.8+4.2, the proto-planetary nebula CRL 618 was also observed on 28 and 29 December 2013. The observing parameters are summarised in Table \ref{crl618}. 

When a source is very weakly polarised ($P_c/I \lesssim 0.5\%$), very small changes to the leakage terms can cause drastic changes to the resulting polarisation position angles across the source.  When we analysed the data for CRL~618, we saw position angles that varied by up to 90$^{\circ}$ from night to night depending on how exactly we reduced the data.  The reason for this was that the source was extremely bright at 1.3\,mm ($\sim$\,2\,Jy/beam), but was very weakly polarised ($\sim$\,0.5\% polarisation at the intensity peak), which caused us to hit a dynamic range limit  where the polarisation detected at very low levels can be caused by imperfections in the leakage solutions.

The scatter in the real and imaginary parts of the leakage solutions on the two nights was $\sim$\,0.01, which is larger than normal considering the 2 consecutive nights period; however, in this case  those slight differences led to significantly different maps for $Q$ and $U$.   These small changes in leakages had caused virtually no variations in the CARMA polarisation maps of protostellar cores (e.g. \citealt{Stephens2013, Hull2013, Hull2014}), which were on average at least a few percent polarised, and which tended to be much fainter sources where detection of polarisation was limited by the system temperature instead of by dynamic range.

We therefore urge caution when interpreting CARMA observations of sources with polarisation fractions of $< 0.5\%$.  While the very low polarisation percentage limit we find with CARMA is consistent with the value of 0.7\% found with the SMA data, we are not able to confidently determine the B-field configuration.
\newpage
\begin{table}
\small\addtolength{\tabcolsep}{-4pt}
\caption[]{\label{crl618} CARMA observations of CRL 618}
%\hspace{1cm}
%\begin{flushleft}
\begin{tabular}{|l|l|l|}
\hline
Source & CRL 618    &  \\
\hline
Phase center & $\alpha$=04:42:53.6 & $\delta$=36:06:53.4 \\
Obs. Date & 2013-12-28   & 2013-12-29  \\
Gain calibrator & 3C111 & 3C111  \\
Passband calibrator & 3C84 & 0510+180 \\
Flux calibrator & 3C84 & Uranus    \\
Total project length (hrs) & 3.8 & 5.5  \\
Time on source (hrs) & 2.43 & 3.28 \\
Total Opacity$^{\dagger}$ & 0.55 & 0.86  \\
\hline
\hline
\end{tabular}
\begin{minipage}{10cm}
\hspace{-1cm}
\hspace{1cm}$\dagger$ Opacity at 230 GHz due to phase noise and atmospheric absorption
\end{minipage}
%\end{flushleft}
\end{table}

\bsp
\label{lastpage}

\end{document}